\documentclass[aps,twocolumn, amsmath,amssymb,pra,superscriptaddress]{revtex4-2}

\usepackage{graphicx}
\usepackage[dvipsnames]{xcolor}
\usepackage{bm}

\usepackage{empheq,etoolbox}
\usepackage[makeroom]{cancel}
\usepackage{bbold}
\usepackage{comment}
\usepackage{appendix}

\usepackage{mathabx}
\usepackage{amsmath}
\usepackage{booktabs}
\usepackage[T1]{fontenc}

\usepackage[section]{placeins}

\renewcommand{\i}{i}

\usepackage{verbatim}
\usepackage{xcolor}
\definecolor{lightgrey}{rgb}{0.8, 0.8, 0.8}
\definecolor{misprint}{rgb}{0, 0, 1}
\definecolor{remark}{rgb}{0.0, 0.5, 0.69}

\definecolor{purple}{rgb}{0.8,0,0.6}
\definecolor{darkgreen}{rgb}{0.00,0.6,0.00}

\begin{document}

\title{Cosmological structure formation in ultralight dark matter}

\author{K. Korshynska}
\affiliation{Fundamentale Physik für Metrologie FPM, Physikalisch-Technische Bundesanstalt PTB, Bundesallee 100, 38116 Braunschweig, Germany}
\affiliation{Institut für Mathematische Physik, Technische Universität Braunschweig, Mendelssohnstraße 3, 38106 Braunschweig, Germany}
\author{A. I. Yakimenko}
\affiliation{Department of Physics, Taras Shevchenko National University of Kyiv,
64/13, Volodymyrska Street, Kyiv 01601, Ukraine}
\author{Y. Revaz}
\affiliation{Laboratoire d’Astrophysique, EPFL, Observatoire de Sauverny, 1290 Versoix, Switzerland}
\author{E. V. Gorbar}
\affiliation{Department of Physics, Taras Shevchenko National University of Kyiv, 
64/13, Volodymyrska Street, Kyiv 01601, Ukraine}
\affiliation{Bogolyubov Institute for Theoretical Physics, 14-b Metrolohichna Street, Kyiv 03143, Ukraine}

\begin{abstract}
An important challenge in cosmology is to describe the formation of galaxies and structures, starting from small inhomogeneities in the early Universe, which later grew and formed the so- called dark matter halos. We consider the dark matter to be composed of ultralight bosons and simplify the general complex problem of halo formation by using the time-dependent variational analysis approach. This yields, for the first time, a fully analytical equation for the time-evolution of perturbations towards isolated haloes. We follow the evolution of the halo size and density during different stages of its formation, revealing the role of the background density, Universe expansion, and quantum pressure.
\end{abstract}

\maketitle

{\it Introduction and model}-- While the existence of dark matter (DM) was first predicted based on astrophysical observations \cite{Zwicky:1933gu,Andernach:2017vpm}, its composition and properties play an important role across various fields of physics, astrophysics, and cosmology. The widely used $\Lambda$CDM model of cosmology assumes that the DM is a non-relativistic collisionless fluid, referred to as cold dark matter (CDM) \cite{blumenthal1984formation}. The CDM explains many of the observed astrophysical phenomena, however encounters a few challenges, such as cusp-core tension \cite{weinberg2015cold, de_Blok_2009}, which could be resolved by accounting for the microscopic nature of DM. This characterizes the ultralight dark matter (ULDM) class of models, which assume that the DM is composed of ultralight bosons in the mass range $m \in [10^{-18}, 10^{-24}]$~eV/$c^2$  \cite{Ferreira,jackson2023search, Schive_2026, chavanis2025review}. The question of viability of the ULDM can be addressed from an astrophysical perspective based on such observations as rotation curves \cite{bar2018galactic, Korshynska_2023}, velocity dispersion \cite{Goldstein, 7fp7-klxf}, and from a cosmological perspective by analyzing the evolution of the DM in an expanding Universe
\cite{binney2011galactic, Marsh_2016, PhysRevD.82.103528}. 

According to the standard $\Lambda$CDM model of cosmology, dark matter comprises about $85\%$ of the matter in the Universe and governs its expansion during the matter-dominated epoch, between redshifts $z_i \cong 3100$  and  $z_f \cong 0.5$ ($t_f = 9.8 \times 10^9$~yr.)  \cite{binney2011galactic, Planck2018}. During this epoch, galaxy formation takes place in the flat Einstein-de Sitter universe \cite{binney2011galactic}; the scale factor evolves as $a(t) = (t/t_f)^{2/3}$ to its final value, which, for convenience, we hereby set to $a(t_f) = 1$. Both CDM and ULDM rely on the cosmological principle that assumes the Universe, and hence the matter is homogeneous and isotropic at very large scale.  The spacetime metric of the homogeneous expanding Universe is given by the Friedmann-Lema\^{\i}tre-Robertson-Walker (FLRW) metric  \cite{binney2011galactic, rubakov2018introduction}, in which the  
bosonic ULDM obeys the Klein-Gordon equation of motion
\begin{eqnarray}
    \ddot{\psi} &+& 3H \dot{\psi} + \left(\frac{mc^2}{\hbar}\right)^2\psi_0 = 0\, , \label{eq: backgroun relativistic}
\end{eqnarray}
where $m$ denotes the DM particle mass and $H = \dot{a}/a$ is a Hubble parameter.
The exact solution to Eq.~(\ref{eq: backgroun relativistic}) exhibits fast oscillations starting at $t \sim H^{-1}$ with period $\hbar/(mc^2)$ \cite{PhysRevD.82.103528, Marsh_2016}. For a typical boson mass range $m = 10^{-22}-10^{-20}$~eV/$c^2$, fast oscillations start well before the matter-radiation equality and have a very small period compared to $H^{-1}$ \cite{PhysRevD.82.103528}. Therefore, for the considered matter-dominated epoch, one can exclude these fast oscillations by taking the nonrelativistic limit $\psi= e^{imc^2 t/\hbar} \psi_0$ in Eq.~(\ref{eq: backgroun relativistic}), which gives
\begin{equation}
     \dot{\psi}_0 + \frac{3}{2}H \psi_0 = 0, 
\label{background-evolution}
\end{equation}
and agrees with the standard result in cosmology for nonrelativistic matter \cite{rubakov2018introduction, binney2011galactic}, implying that
\begin{equation}
     |\psi_0|^2 = \rho_0 = \frac{3}{8\pi G} \left(\frac{\dot{a}}{a}\right)^2 = \frac{1}{6\pi G t^2}\, .
     \label{eq: background dens}
\end{equation}
Hence, ULDM in a non-relativistic limit and the cosmological setting has the same behavior as $\Lambda$CDM \cite{PhysRevD.82.103528} reproducing the observed distribution of matter at large scales \cite{schive2014cosmic}.

While the cosmological principle holds on large scales, on the smaller scales the Universe shows inhomogeneities such as galaxies and clusters of galaxies \cite{binney2011galactic}.  To describe the formation of these structures, one considers small density perturbations on the uniform density (\ref{eq: background dens}), see Ref.~\cite{harko2011evolution}, which generate small metric perturbations described by the gravitational potential $\Phi(\mathbf{r},t)$. The metric for a scalar perturbation can be written in the linearized pseudo-Newtonian form \cite{PhysRevD.48.4623}
\begin{equation}
    ds^2 = \left(1 + \frac{2\Phi}{c^2}\right) c^2 dt^2 + a^2(t) \left(-1 + \frac{2\Phi}{c^2}\right) \delta_{ij} dx^i dx^j\, ,
    \label{eq: perturbed FLRW metric}
\end{equation} 
where $\Phi/c^2 \ll 1$, see Refs.~\cite{Hwang_2008, PhysRevD.107.083508} for further discussion of the post-Newtonian corrections. The bosonic field $\psi$ of ULDM in this spacetime obeys the system of Schr\"odinger-Poisson equations (SPE) in the expanding Universe
\begin{subequations}
\label{eq: GPPE system}
    \begin{eqnarray}
    i \hbar \left(\frac{\partial}{\partial t} + \frac{3}{2}H\right)\psi &=& -\frac{\hbar^2}{2ma^2} \nabla^2 \psi + m\Phi\psi\, ,\\ \label{eq: GPE derived}
    \Delta \Phi &=& 4\pi G a^2 (|\psi|^2 - \rho_0)\, ,
\end{eqnarray}
\end{subequations}
as has been derived and extensively analyzed in Refs.~\cite{Widrow:1993qq, Hu_2000}. These equations can be further reformulated in a fluid description using the Madelung transformation \cite{madelung1927z} of the complex field $\psi = \sqrt{\rho}e^{i\theta}$ in terms of its absolute value $\sqrt{\rho(\mathbf{r},t)}$ and phase $\theta(\mathbf{r},t)$.  This leads to the hydrodynamic equations for density $\rho(\mathbf{r})$ and velocity $\mathbf{v} = \hbar/(ma) \cdot \nabla \theta$
\begin{subequations}
\label{eq: hydrodynamic eqs}
    \begin{eqnarray}
    &&\dot{\rho} + 3H \rho + \frac{1}{a} \nabla (\rho \mathbf{v}) = 0,\\
    &&\dot{\mathbf{v}} + H \mathbf{v} + \frac{1}{a} (\mathbf{v} \nabla) \mathbf{v} = - \frac{1}{a} \nabla \Phi - \frac{1}{m a^3} \nabla Q, \\
    &&\Delta \Phi = 4\pi G a^2 (\rho - \rho_0)\, .
\end{eqnarray}
\end{subequations}
{\it Jeans instability and ansatz}-- Without the second term on the right-hand side of Eq.~(\ref{eq: hydrodynamic eqs}b), those equations are the equivalent of the continuity equation for a collisionless (pressureless) fluid, corresponding to CDM. The additional term describes the quantum pressure potential $Q = -\hbar^2/(2m) \cdot (\Delta \sqrt{\rho}) / \sqrt{\rho}$ and is a unique feature of ULDM compared to CDM, producing repulsion to counteract attractive gravity \cite{chavanis2019predictive}. The balance between them is reached for 
ULDM overdensities with momentum 
\begin{eqnarray}
    k_J = \left( \frac{16\pi G \rho_0 m^2 a^4}{\hbar^2}\right)^{1/4} \label{eq: Jeans wavenumber}\,,
\end{eqnarray}
known as the critical Jeans wavenumber in the expanding Universe \cite{chavanis2012growth} (see also Refs.~\cite{ourabah2022gravitational,suarez2018jeans, Chavanis_2021} and Supplemental Material~\cite{supp}). Due to the Jeans instability \cite{Hu_2000, Harko_2019, suarez2018jeans, ourabah2022gravitational, Chavanis_2021}, density fluctuations with $k < k_J$ grow and undergo gravitational collapse. Thus, in terms of the Jeans wavelength $\lambda_J = 2\pi/k_J$, perturbations with length scales larger than $\lambda_J$ cause instability and may lead to structure formation \cite{Hu_2000} with the corresponding Jeans mass \cite{Harko_2019, ourabah2022gravitational,suarez2018jeans}
\begin{equation}
    M_J = \frac{4 a^3}{3}\pi \rho_0 \left(\frac{\lambda_J}{2}\right)^3 =  \frac{\pi^4}{6} \frac{\hbar^2}{Gm^2 \lambda_J} \label{eq: Jeans mass 1}\, .
\end{equation}
Notably, the length and mass scales $\lambda_J$ and $M_J$ are finite and determine the lower limit of the size and mass of ULDM halos formed due to the gravitational collapse of a homogeneous distribution of non-interacting bosons \cite{chavanis2011mass}.
It is worthwhile to remind here that in CDM (pressureless fluid) corresponding to $\hbar = 0$ and $M_J = 0$, density inhomogeneities are unstable on all scales inside the Hubble radius since CDM has no quantum pressure.

While from the discussion above, we can get some qualitative insight into structure formation in ULDM, the complete description of this process in the expanding Universe is computationally demanding and requires large-scale numerical simulations of the Schr\"odinger-Poisson equations \cite{Schive_2026, schive2014cosmic, 10.1093/mnras/stab1764, Mina_2022, Veltmaat,Nori,Mocz}. 
While these simulations provide a final state of an ULDM halo for a specific given set of system parameters, there have been many efforts to understand its structure while covering the whole parameter space with analytical approximations and variational methods \cite{chavanis2011mass, chavanis2019predictive, Goodman_2000, schive2014understanding}. 
In this Letter, we extend these methods to the cosmological problem of structure formation, providing analytical insights that complement and help validate previous numerical studies.

In what follows, we consider an overdensity $\delta$ with respect to the background density
\begin{subequations}
\label{eq:coherent wavefunction ansatz}
    \begin{eqnarray}
    \rho(\mathbf{r},t) &=& \rho_0(t)(1 + \delta(\mathbf{r},t))\\
    \psi(\mathbf{r},t) &=& \sqrt{\rho(\mathbf{r},t)}e^{i\theta(\mathbf{r},t)}\, ,
\end{eqnarray}
\end{subequations}
where $\rho_{0}$ is the background density in the matter-dominated FLRW spacetime given by Eq.~(\ref{eq: background dens}), c.f., Ref.~\cite{rubakov2018introduction}. A simple model to describe the formation of a galaxy would be to consider a spherically symmetric overdensity of a shape
\begin{equation}
    \delta(r, t) = \frac{R_0^3}{R(t)^3} e^{-r^2/R(t)^2} \label{eq: delta}\, ,
\end{equation}
where $R(t)$ is the size of the halo. Here, the distance scale $R_0$ is related to the total mass of the ULDM inhomogeneity $M = \int |\psi|^2 d\mathbf{r}= \pi^{3/2} R_0^3 \rho_f$. We note that this type of ansatz was first introduced to describe nonlinear pulse propagation in optical fibers \cite{PhysRevA.27.3135}, an atom BEC in a harmonic trap \cite{SALASNICH_2000}, and then further developed in the context of ULDM in the absence or weak ULDM self-interaction \cite{chavanis2011mass}. Armed with ansatz (\ref{eq:coherent wavefunction ansatz}), one maps the problem of solving the SPE system (\ref{eq: GPPE system}) for field $\psi(\mathbf{r}, t)$ to a simpler problem with time-dependent parameter $R(t)$. In the discussed cosmological setting, we expect that the perturbation $\delta$ is small in the linear regime realized initially and becomes locally much more dense than the background later in the nonlinear regime.  The density inhomogeneity $\rho(\mathbf{r}) - \rho_0$ generates the gravitational potential $\Phi$ according to Eq.~(\ref{eq: GPE derived})
\begin{eqnarray}
    \Phi =  - G a^2 \rho_0\pi^{3/2} R_0^3 \frac{\textrm{Erf}[r/R]}{r}\, \label{eq: phi},
\end{eqnarray}
where the contribution from the background density $\rho_0$ is given by $\Phi_0 = G a^2 \rho_0\frac{2}{3}\pi r^2$ and is subtracted from the total gravitational potential to obtain the gravitational potential $\Phi$ of the perturbation.

The corresponding ansatz for the phase $\theta(\mathbf{r},t)$ of the wavefunction (\ref{eq:coherent wavefunction ansatz}b) has to satisfy the continuity equation 
(\ref{eq: hydrodynamic eqs}a). For the considered density ansatz (\ref{eq:coherent wavefunction ansatz}a) this equation reads
\begin{equation}
    \dot{\delta} + \frac{1}{a} \nabla \left[(1 + \delta) \mathbf{v}\right] = 0\, . \label{eq: cont with delta}
\end{equation}
In accord with the spherical symmetry of our variational ansatz we assume that $\mathbf{v} = v(r)\mathbf{e}_r$ is directed along $\nabla \delta$. 
In addition, we expect that the physically meaningful solution should be finite at $\mathbf{r} = 0$. Thus, we find
\begin{subequations}
\label{eq: vj(r)}
    \begin{eqnarray}
    v(r) &=& \frac{r a(t)}{1 + e^{r^2/R^2}(R/R_0)^3} \frac{\dot{R}}{R} =  \frac{r a \delta}{1 + \delta}  \frac{\dot{R}}{R} \label{eq: v(r)}\\
    \mathbf{j}(r) &=& \rho(r)v(r) \mathbf{e}_r = \rho_0a \frac{R_0^3}{R^3} e^{-r^2/R^2} r   \frac{\dot{R}}{R} \mathbf{e}_r\, \label{eq: j(r)},
\end{eqnarray}
\end{subequations}
so that $\mathbf{v} = 0$ in equilibrium when $\dot{R} = 0$.
The result (\ref{eq: vj(r)}) implies that the current $\mathbf{j}$ and the velocity $v(r)$ decay exponentially at large distances from the fluctuation. At small distances, in contrast, when the fluctuation density dominates over the background, i.e., $R \ll R_0$, we recover the velocity profile $\mathbf{v} = a(t) \frac{\dot{R}}{R} r \mathbf{e}_r$ and the corresponding phase $\theta \sim r^2$ as is well known for the variational analysis of the Gaussian density in the absence of background, c.f. Refs.~\cite{salasnich2000time} and further details in Supplemental Material~\cite{supp}. In the general case, phase $\theta(\mathbf{r}, t)$ can be derived from Eq.~(\ref{eq: v(r)}) and reads
\begin{equation}
    \theta(r, t) =  - \beta R^2 \textrm{ln} \left(1 + \left(\frac{R_0}{R}\right)^3 e^{-r^2/R^2}\right)\, ,\label{eq: phase ansatz}
\end{equation}
where $\beta(t) = ma^2/(2\hbar) \cdot \dot{R}/R$. Following the ideas of Refs.~\cite{SALASNICH_2000, chavanis2016collapse}, we introduce $\beta(t)$ as a canonically conjugate variational parameter to the halo size $R(t)$. Here, this particular parametrization is chosen to be analogous to the previous studies with a Gaussian ansatz in the absence of background \cite{SALASNICH_2000}, when $\theta(r, t) = \beta(t) r^2$.

{\it Equations of motion}-- Let us now determine the dynamics of the ULDM following the approach of Refs.~\cite{SALASNICH_2000, chavanis2011mass} using the SP action
\begin{eqnarray}
    S_\textrm{SP} = \int dt \mathcal{L} &=& \frac{a^3}{m}\int dt \int d\mathbf{r} \psi^* \left[i \hbar \frac{\partial}{\partial t} + \frac{3i\hbar}{2}H  \right. \nonumber\\
    &&+ \left. \frac{\hbar^2}{2ma^2} \nabla^2 - \frac{m}{2}\Phi \right] \psi\, \label{eq: SGPP}.
\end{eqnarray}
Here we plug our ansatz (\ref{eq:coherent wavefunction ansatz}) for the wavefunction $\psi$ using the Gaussian ansatz (\ref{eq: delta}) for overdensity $\delta$, Eq.~(\ref{eq: phase ansatz}) for phase $\theta$, and 
Eq.~(\ref{eq: phi}) for gravitational potential $\Phi$.
With this, we find 
\begin{eqnarray}
    &&\mathcal{L} = \frac{2 \pi \hbar \rho_0 a^3}{m} \left(2R  \left(\dot{\beta}R + 2 \beta \dot{R}\right) \int dr r^2 (1 + \delta) \log \left[1 + \delta\right] \right. \nonumber \\
     &&-\left. \frac{\hbar}{ma^2}\left[4\beta^2 + \frac{1}{R^4} \right] \int dr r^4 \frac{\delta^2}{1 + \delta} + \frac{\pi^{3/2} G ma^2 \rho_0 R_0^6}{2^{3/2} \hbar R}\right)\,, \label{eq: lagrangian}
\end{eqnarray}
see details on the derivation in Supplemental Material~\cite{supp}.
Here the first line describes the time dependence of the wavefunction parameterized by the time derivative of the phase $\theta$. The first term in the second line is the kinetic energy -  both classical $\sim (\theta')^2 \sim |v|^2$ and quantum, associated with the quantum pressure, c.f. Refs.~\cite{chavanis2019predictive, Liu_2023}. The last term is the gravitational energy due to the potential $\Phi$ generated by the fluctuation. The common multiplier of the Lagrangian (\ref{eq: lagrangian}) is constant since $a^3 \sim t^2$ and $\rho_0 \sim 1/t^2$ (see Eq.~(\ref{eq: background dens})).

For the given Lagrangian (\ref{eq: lagrangian}) the evolution of two variational parameters $R(t)$ and $\beta(t)$ can be predicted from the Lagrange equations of motion. The equation for $\beta(t)$ simplifies to the solution of the continuity equation $\beta(t) = ma^2/(2\hbar) \cdot \dot{R}/R$, as shown in Supplemental Material~\cite{supp}. The Lagrange equation of motion for $R(t)$ reads
\begin{subequations}
\label{eq: for R(t)}
    \begin{eqnarray}
    &&\ddot{R} + 2\frac{\dot{a}}{a} \dot{R} + \frac{\dot{R}^2}{R} \left(\frac{I_2}{2I_1} -1\right) + \frac{3 \sqrt{\pi}}{32 \sqrt{2} I_1} \left(\frac{\dot{a}}{a}\right)^2 R \nonumber\\
    &&+ \frac{\hbar^2 }{m^2 a^4 R^{3}}\left(\frac{I_2}{2I_1} - 2\right) = 0, \label{eq: ddotR}\\
    &&I_1(R) = \int_0^{+\infty} dx \frac{x^4 e^{-2x^2}}{1 + (R_0/R)^3e^{-x^2}}, \label{eq: I1R}\\
    &&I_2(R) = \int_0^{+\infty} dx \frac{x^4 (2x^2 - 3) e^{-2x^2}}{1 + (R_0/R)^3e^{-x^2}}, \label{eq: I2R}
\end{eqnarray}
\end{subequations}
where we introduced a dimensionless integration variable $x = r/R$ and used $\rho_0 = 3/(8\pi G) \cdot (\dot{a}/a)^2$ from Eq.~(\ref{eq: background dens}). Integrals $I_{1/2}(R)$, which quantify different contributions in Eq.~(\ref{eq: ddotR}), are illustrated and further discussed in Supplemental Material~\cite{supp}. 

{\it Linear and nonlinear regimes}-- Having found the equation of motion for the halo size in the expanding Universe, we in what follows investigate the two limits, frequently examined in the literature, where the halo density is either small (linear stage) or locally dominant (nonlinear stage) compared with the background density. During the linear stage $(R_0/R)^3 \ll 1$, we have $I_1 = 3\sqrt{\pi}/(32\sqrt{2})$ and $I_2 = -I_1/2$. Hence, the equation for $R(t)$ simplifies to
\begin{eqnarray}
    \ddot{R} + 2\frac{\dot{a}}{a} \dot{R} -\frac{5}{4}\frac{\dot{R}^2}{R}  + \left(\frac{\dot{a}}{a}\right)^2R - \frac{9\hbar^2 }{4m^2 a^4 R^{3}} = 0\, , \label{eq: linear regime}
\end{eqnarray}
where the fourth term is the attractive gravity and the last term is the repulsive quantum pressure. We find that these two effects compensate each other at the critical fluctuation size
\begin{equation}
    R_J = \left(\frac{a}{\dot{a}}\right)^{1/2} \left(\frac{9\hbar^2}{4m^2 a^4}\right)^{1/4}\, \label{eq: RJ} ,
\end{equation}
which, up to a prefactor $R_J/\lambda_J \approx 0.3$, coincides with the Jeans wavelength $\lambda_J = 2\pi/k_J$, where $k_J$ is given by Eq.~(\ref{eq: Jeans wavenumber}) and relation (\ref{eq: background dens}) is used. Thus, in accord with the Jeans instability mechanism, for an initial condition $R> R_J$ and $\dot{R} = 0$ the linear gravitational attraction $\sim R$ dominates and drives the collapse of the halo.
In the ultimate case of negligible quantum pressure, Eq.~(\ref{eq: linear regime}) can be solved assuming $R = A t^\alpha$, which leads to
the two solutions $\alpha=2(1\pm\sqrt{5})/3$, with the negative branch describing contraction of the overdensity. The corresponding central contrast evolves as $\delta(0,t)\sim t^{2(\sqrt{5}-1)}$, rather than the standard pressureless linear growing mode $\delta \sim t^{2/3}$ \cite{chavanis2012growth}. This discrepancy reflects the imposed relation between the amplitude and width of the Gaussian profile (\ref{eq: delta}).



In the nonlinear regime $(R_0/R)^3 \gg 1$, we have $I_1 = 3\sqrt{\pi}/8 \times (R/R_0)^3$ and $I_2  = 2I_1$ that gives
\begin{eqnarray}
    \ddot{R} + 2\frac{\dot{a}}{a} \dot{R} - \frac{\hbar^2}{m^2 a^4 R^{3}}  + \frac{1}{4 \sqrt{2}}\left(\frac{\dot{a}}{a}\right)^2 \frac{R_0^3}{R^2}  = 0\, \label{eq: nonlin R eq}.
\end{eqnarray}
The latter equation unlike Eq.~(\ref{eq: linear regime}), where the gravitational term is $\sim R$, includes gravity as a nonlinear term $\sim 1/R^3$. This behavior allows the gravitational attraction to play an important role throughout the entire evolution of a ULDM halo. We also note that Eq.~(\ref{eq: nonlin R eq}) can be obtained as an exact Lagrange equation for the parameter $R(t)$ if, instead of wavefunction (\ref{eq:coherent wavefunction ansatz}), one would consider 
\begin{equation}
    \psi = \sqrt{\rho_0}(e^{i\alpha} + \sqrt{\delta}e^{i\theta}) \label{eq: incoherent psi}
\end{equation}
in the action of SP equations given by Eq.~(\ref{eq: SGPP}). Here phase $\alpha(\mathbf{r})$ is a coordinate-dependent random phase, which makes the background and the fluctuation an incoherent mixture, similarly to the description of gravitating solitons and finite-temperature BECs in Refs.~\cite{PhysRevLett.127.014101, Griffin_Nikuni_Zaremba_2009}. This reflects halo-background decoupling rather than dephasing because, as soon as the background amplitude becomes negligible, the relative phase becomes dynamically irrelevant. At this point an isolated halo is formed, see further details in Supplemental Material~\cite{supp}.

{\it Numerical analysis}-- To describe the complete evolution of the halo size, we solve Eq.~(\ref{eq: for R(t)}) numerically. For that we account for the time dependency of scale factor $a(t) = (t/t_f)^{2/3}$ and introduce convenient dimensionless variables $y = R/R_0$ and $\tau = t/t_f$, in terms of which the equation reads
\begin{subequations}
\label{eg for y(tau)}
    \begin{eqnarray}
    &&\frac{d^2 y}{d\tau^2} + \frac{4}{3\tau} \frac{dy}{d\tau} + \frac{1}{y} \left(\frac{dy}{d\tau}\right)^2 \left(\frac{I_2}{2I_1} -1\right) + \frac{\sqrt{2\pi} y}{48 I_1 \tau^2} \nonumber\\
    &&+ \frac{q}{\tau^{8/3} y^3}\left(\frac{I_2}{2I_1} - 2\right) = 0,\\
    &&I_1(y) = \int_0^{+\infty} dx \frac{x^4 e^{-2x^2}}{1 + e^{-x^2}/y^3},\\
    &&I_2(y) = \int_0^{+\infty} dx \frac{x^4 (2x^2 - 3) e^{-2x^2}}{1 + e^{-x^2}/y^3},
\end{eqnarray}
\end{subequations}
where we used $q = \hbar^2 t_f^2/(m^2 R_0^4)$ as a dimensionless parameter characterizing quantum pressure. Here the length scale $R_0 = (M/(\pi^{3/2} \rho_f))^{1/3}$ is related to the total mass $M$ of overdensity (\ref{eq: delta}). 

To fully formulate the problem, we need to fix our initial conditions for the density contrast $\delta(t_i)$ of
the primordial fluctuations
that will lead to a halo of given mass.
For that, we take input from the cosmic microwave background (CMB) that was released at the epoch of recombination, approximately $t_i = 3.8 \times 10^{5}$ years ($z = 1100$) after the Big Bang. Measurements of the CMB temperature and polarization anisotropies show that the Universe contained small primordial fluctuations, with a characteristic amplitude of order $\Delta T/T\sim10^{-5}$ \cite{Planck2018}. These temperature fluctuations allow reconstructing a CDM total matter power spectrum at $z=1100$ using the CAMB code
\cite{lewis2026}, assuming a \cite{Planck2018} cosmology.
For a given ULDM particle mass, the power spectrum is then modified using the corresponding
ULDM transfer function \cite{hu2000}.
To relate the final halo mass and the density contrast, we first compute the
value of the perturbation wavenumber at the origin of collapse. This is done
by considering the mass enclosed in a sphere of radius equal to half its
wavelength \citep[e.g.]{bullock2017}.  Having determined $k$, the density
contrast is obtained from the power spectrum at this wavenumber.

\begin{figure}[htp]
    \centering  \includegraphics[width = 0.5\textwidth]{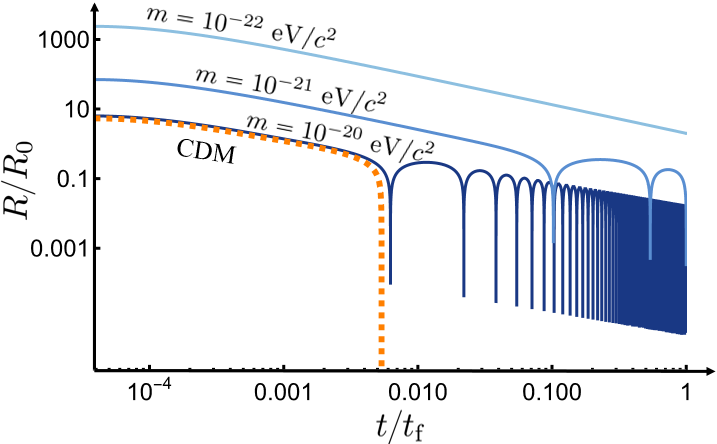}
    \caption{Numerical solution to Eq.~(\ref{eg for y(tau)}) assuming the initial condition $dy(\tau_i)/d\tau = \dot{R}(t_i) = 0$, $y(\tau_i) = R(t_i)/R_0 = \delta^{-3}(\tau_i)$ at the initial point of time $\tau_i = 3.88 \times 10^{-5}$ ($t_i = 3.8 \times 10^{5}$~\textrm{yr}). Here $\delta(\tau_i)$ is evaluated with the CAMB code and equals $7.372 \times 10^{-11}$, $2.864 \times 10^{-6}$, and $3.944 \times 10^{-3}$ for the boson masses $10^{-22}$~eV/c$^2$, $10^{-21}$~eV/c$^2$, and $10^{-20}$~eV/c$^2$ correspondingly. The solution $R(t)$ for these different values of the boson mass $m$ is shown with solid blue lines. For the CDM $\delta(\tau_i) = 6.335 \times 10^{-3}$ and $q=0$, see orange dashed line. For all lines, the halo mass $M = 10^8 M_\odot$ is fixed, giving $R_0 = 52.7$~kpc. Note that while the time-scale of the oscillations for $\tau > 10^{-2}$ appears to decrease on the log-scale, it remains constant on a linear scale, c.f., Fig.~\ref{fig: Energy plot}.}
    \label{fig: R solution}
\end{figure}

The primordial fluctuations provide the initial conditions for the subsequent growth of cosmic structure through gravitational instability, as shown in Fig.~\ref{fig: R solution}. A CDM halo experiences singular collapse due to the absence of quantum pressure to compensate gravity, resulting in zero Jeans scale. On the other hand, a ULDM halo collapses due to Jeans instability $y(\tau_i) < R_J/R_0$, see Eq.~(\ref{eq: RJ}), and further enters the nonlinear stage only if the initial overdensity $\delta(\tau_i) = y^{-3}(\tau_i)$ fulfills $\delta(\tau_i) > 1.5 \times 10^{-8}$. This limits the boson mass $m$ from below, since for small boson masses with $\delta(\tau_i) < 1.5 \times 10^{-8}$ the collapse within the matter-dominated epoch is not possible, see $m = 10^{-22}$~eV/c$^2$ in Fig.~\ref{fig: R solution}.

\begin{figure}[htp]
    \centering  \includegraphics[width = 0.49\textwidth]{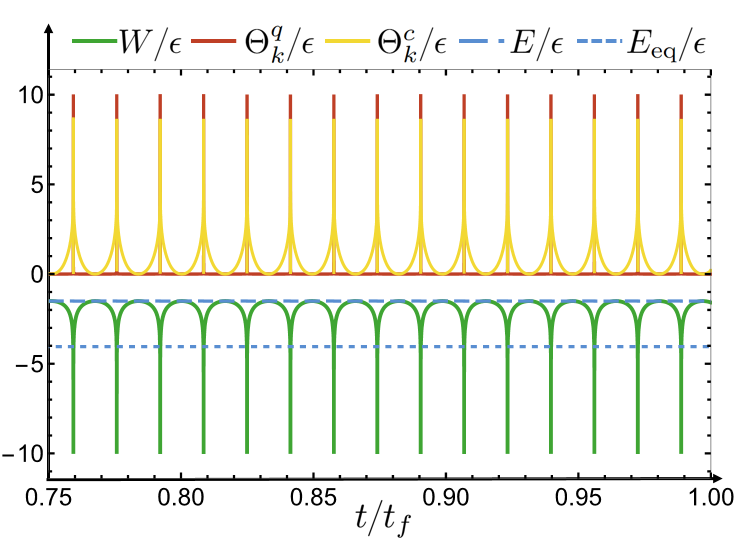}
    \caption{Numerical solution for the quantum kinetic energy $\Theta_k^q$, classical kinetic energy $\Theta_k^c$, gravitational energy $W$ and total energy $E = \Theta_k^q + \Theta_k^c + W$ depending on $\tau = t/t_f$ in the interval $\tau \in [0.75, 1]$ as given by Eq.~(\ref{eq: energy components}) for $R(t)$ presented in Fig.~\ref{fig: R solution}. The line $E_\textrm{eq}$ stands for the stationary solution at $y_\textrm{eq} = 9\sqrt{2} q$ in the case of a ULDM halo decoupled from the background. Here all energies are given in the units of $\epsilon = R_0^5/(Gt_f^4)$ and the boson and halo masses are fixed to $m = 10^{-20}$~eV/c$^2$ and $M = 10^8 M_\odot$. The $y$-axis is scaled on a symmetric arcsinh-scale.}
    \label{fig: Energy plot}
\end{figure}

We also find that after a ULDM halo decouples from the background density, its size $R(t)$ oscillates. To further investigate this dynamics, presented in Fig.~\ref{fig: R solution}, we analyze the energy of the system. The kinetic energy $\Theta_k$  and gravitational energy $W$, which enter Lagrangian (\ref{eq: lagrangian}), read 
\begin{subequations}
\label{eq: energy components}
    \begin{eqnarray}
    &&\Theta_k = \Theta_k^q + \Theta_k^c,\\
    &&\Theta_k^q = \frac{R_0^5}{3G t_f^4} \frac{I_1(y)}{y^3} \frac{q}{\tau^{4/3} y^2},\\
     &&\Theta_k^c = \frac{R_0^5}{3G t_f^4} \frac{I_1(y)}{y^3} \tau^{4/3} \left(\frac{dy}{d\tau}\right)^2,\\
     &&W = -\frac{\sqrt{\pi}}{36\sqrt{2}} \frac{R_0^5}{G t_f^4} \frac{1}{\tau^{2/3}y},
\end{eqnarray}
\end{subequations}
where the dimensionless units $\tau = t/t_f$ and $y = R/R_0$ are used, see details of the derivation in Supplemental Material~\cite{supp}. Here we explicitly separated the classical kinetic energy $\Theta_k^c$ due to the velocity $dy/d\tau$ of the change of the halo size and the quantum kinetic energy $\Theta_k^q \sim q$, which is induced by the quantum pressure. The latter is a unique feature of ULDM models, c.f. Refs.~\cite{chavanis2019predictive, Liu_2023} and is sometimes referred to as thermal energy \cite{holland1995quantum}. The three energy components are illustrated in Fig.~\ref{fig: Energy plot}, from where we also see that the total energy $E = \Theta_k + W$ of the ULDM halo becomes time-independent during the nonlinear stage, when the halo decouples from the background. 

At the minima of $R(t)$ both the kinetic and the gravitational energy have sharp extrema, and further gravitational collapse is prevented mainly by quantum pressure, which dominates the kinetic energy. Thus, our findings show how the quantum pressure stabilizes a ULDM halo and sets a finite minimum length scale of the structure formation. At plateaus between the minima of $R(t)$, all energy components approach their absolute minimum values corresponding to the big halo size $R(t)$ and small velocity $\dot{R}(t)$. At these points, the static equilibrium is, however, not established due to the presence of an excess energy in the system compared with the energy \begin{equation}
    E_\textrm{eq} =  - \frac{\sqrt{\pi}}{1296 q} \frac{R_0^5}{G t_f^4}
\end{equation}
of the static solution of the same mass
\begin{equation}
       R_\textrm{eq} = \frac{3}{\sqrt{2}\pi} \frac{\hbar^2}{ G \rho_f R_0^3 m^2} = 0.99 \left(\frac{10^{-22}~\textrm{eV}/c^2}{m}\right)^{2/3} \textrm{kpc}, \label{eq: Req}\\
\end{equation}
decoupled from the background and the Universe expansion, see Supplemental Material~\cite{supp} for details. In the given dimensionless units, it reads $y_\textrm{eq} = R_\textrm{eq}/R_0 = 9\sqrt{2}q$. Since the total energy of the halo $E$ is larger than the energy $E_\textrm{eq}$ of an equilibrium Gaussian soliton of the same mass, the isolated halo cannot evolve into such a soliton but instead evolves into a higher energy state.

\begin{figure}[htp]
    \centering  \includegraphics[width = 0.5\textwidth]{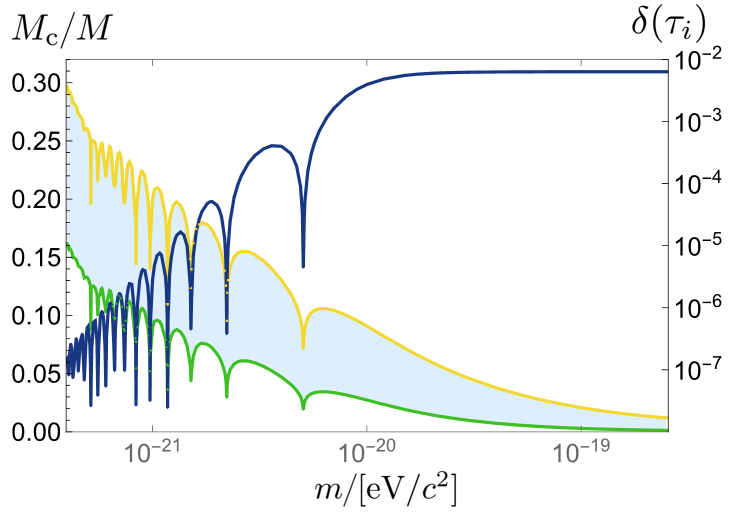}
    \caption{Ratio $M_\textrm{c}/M$ of the core and total halo masses (left $y$-axis, linear scale) and the initial overdensity $\delta(\tau_i) = y^{-3}(\tau_i)$ used for numerical solution of Eq.~(\ref{eg for y(tau)}) (right $y$-axis, log-scale) depending on the particle mass $m$ for a fixed halo mass $M = 10^8 M_\odot$. The yellow line is calculated assuming energy equipartition $E_\textrm{c}/M_\textrm{c} = E/M$, while the green line corresponds to $E_\textrm{c} = E$, i.e., all energy located in the central core. The dark blue line shows the initial overdensity $\delta(\tau_i)$. The shaded region between the two lines covers all possible intermediate virialization stages.}
    \label{fig: core-halo plot}
\end{figure}


{\it Core and {envelope}--} The density profile of this higher energy state appears more complex than the one introduced by Eq.~(\ref{eq: delta}) due to nonspherical evolution, and hence cannot be described by a spherically-symmetric analytical model, presented in this Letter. This state is formed via the complex virialization dynamics that leads to the formation of a halo composed of a soliton and a Navarro-Frenk-White envelope, which emerges in numerical simulations of ULDM structure formation \cite{Schive_2014-core-halo,May,Liu_2023}. There,  non-spherical gravitational collapse converts potential energy into kinetic (wave) energy. Violent relaxation and wave interference redistribute this energy, while gravitational cooling can transfer energy out of the bound region through the emission of waves, allowing the remaining system to approach virial equilibrium \cite{Seidel_1994, Veltmaat2018}. In the inner region, wave dynamics subsequently leads to the formation of a dense, approximately stationary solitonic core surrounded by an extended virialized envelope \cite{Schive2014, Schwabe2016, Veltmaat2018}. 

While the spherically-symmetric analytical ansatz (\ref{eq: delta}) does not allow for a full description of this virialization dynamics, we can still infer the properties of the final core-envelope configuration from our calculations, using the core-halo relations investigated in Refs.~\cite{bar2018galactic,mocz2017galaxy}. They relate the core energy $E_\textrm{c}$ and mass $M_\textrm{c}$ with the total energy $E$ and mass $M$ of the halo, consisting of the core and an envelope. In accord with them, the fraction of the mass in the soliton core lies between the limit of energy equipartition \cite{bar2018galactic} 
\begin{equation}
    \frac{E_\textrm{c}}{M_\textrm{c}} = \frac{E}{M}\, \label{eq: en equi},
\end{equation}
describing a halo directly after virialization, and the limit of all energy being located in the central core \cite{mocz2017galaxy} 
\begin{equation}
    E_\textrm{c} = E\, \label{eq: Mocz},
\end{equation}
describing a halo at later times after gravitational cooling. We obtain the total energy $E = \Theta_k + W$ for given halo mass $M$ and boson mass $m$ by substituting the numerical solution of Eqs.~(\ref{eg for y(tau)}) into Eqs.~(\ref{eq: energy components}) at a nonlinear stage, when the energy approaches a constant, as illustrated in Fig.~\ref{fig: Energy plot}. The density of the virialized core can be approximated with a Gaussian soliton, see Supplemental Material~\cite{supp}, whose energy at equilibrium reads
\begin{equation}
    E_\textrm{c} = -\frac{G^2 m^2 M_\textrm{c}^3}{6 \pi \hbar^2}\, \label{eq: en to soliton} ,
\end{equation}
and hence the core mass $M_\textrm{c}$ remains the only unknown parameter to be determined from the relations (\ref{eq: en equi}) and (\ref{eq: Mocz}). The core-halo mass ratio $M_\textrm{c}/M$ for these two relations is illustrated in Fig.~\ref{fig: core-halo plot}. We find that, with an increase of the boson mass $m$, the core in the center becomes smaller and the halo tends to the CDM limit with no core. This behavior emerges due to the quantum pressure, whose impact is most pronounced for small $m$. It has a two-fold effect in our calculations: it suppresses initial overdensities, see CAMB code \cite{lewis2026} and Fig.~\ref{fig: R solution}, and impacts further evolution of the halo size, entering Eqs.~(\ref{eg for y(tau)}) with the parameter $q \sim 1/m^2$. In the ultimate case of $m > 10^{-19}$~eV, the core mass constitutes only $\lessapprox 2 \%$ of the total halo mass, and hence the ULDM tends to the CDM limit with $q=0$, i.e., no quantum pressure.

Let us discuss the decrease of the core mass $M_\textrm{c}$ as a function of $m$, predicted by our model, in the context of astrophysical observations. The total halo mass of the order of $10^8 M_\odot$ is typical for the dwarf spheroidal galaxies, orbiting the Milky Way \cite{walker2007velocity}. The stellar dynamics in these galaxies was fitted assuming CDM with a cuspy NFW density distribution \cite{walker2007velocity}, ULDM with a boson mass $\sim 10^{-20}$~eV/$c^2$ \cite{Goldstein}, and $\sim 10^{-22}$~eV/$c^2$ \cite{7fp7-klxf}. It was found that for a bigger boson mass, the central core constitutes only a small fraction of the total halo mass and size, while for the boson mass as small as $\sim 10^{-22}$~eV/$c^2$ the stellar dynamics can be completely described by a Gaussian soliton, hence with only a smaller fraction of halo mass located in the envelope. This example of astrophysical observations shows the same tendency as our results in Fig.~\ref{fig: core-halo plot}, however for a more general conclusion further studies are required.

We also note that the prediction of mass distribution between the core and envelope in a ULDM halo, illustrated in Fig.~\ref{fig: core-halo plot}, could be used to further test the underlying assumptions of the model. For instance, deviations from the results presented in Fig.~\ref{fig: core-halo plot} may indicate that the collapse of a particular halo was non-spherical or significantly influenced by collisions with other haloes during nonlinear stage.

{\it Summary and discussion}-- 
To summarize, in this Letter the formation of a ULDM halo in the expanding matter-dominated Universe was investigated by using the approach of variational ansatz, which complements the previous research on structure formation with numerical simulations \cite{Schive_2026, schive2014cosmic, 10.1093/mnras/stab1764, Mina_2022}. 
This approach allowed us to formulate an analytical expression for the ULDM wavefunction consisting of a uniform background and an inhomogeneous spherically symmetric halo. Making no further assumptions about the relative densities of the halo and the background, we formulated an analytical model, which addresses both linear and nonlinear stages of structure formation. In particular, we derived an effective Lagrangian as a function of two variational parameters characterizing the size and phase of the ULDM halo; in view of the Universe expansion, the Lagrangian is explicitly time-dependent. With this, we isolated different contributions to the equation of motion for the halo size $R(t)$, i.e., gravitational attraction, quantum pressure, classical kinetic energy, and Universe expansion.

The initial conditions for the structure formation are set by the power spectrum of primordial fluctuations, which is sensitive to the quantum pressure. While for a large boson mass $m \gtrsim 10^{-19}$~eV/c$^2$ the initial overdensity is the same as for CDM, for smaller boson masses the quantum pressure becomes significant and suppresses density fluctuations, providing much smaller initial overdensities. For the subsequent evolution, during the linear stage when the density fluctuations are small, the coherence between them and the uniform background plays a crucial role in the structure formation.
According to our results, this leads to a gravitational attraction linear in $R(t)$ and drives the gravitational collapse towards a $\sim 10^3$ times more compact halo compared with the initial small overdensity.

The presence of a uniform background plays a significant role only during the linear stage of structure formation, while it can be neglected during the nonlinear stage when the halo completely decouples from the background. Moreover, in contrast to the linear stage, the gravitational term in the equation for the halo size evolution is nonlinear $\sim 1/R^2$, as also evident from the variational analysis of a Gaussian halo density in a static Universe \cite{chavanis2011mass}. The self-gravity is counteracted by the quantum pressure, which is a unique feature of ULDM and suppresses structure formation on small scales. The interplay of these two effects leads to oscillations of the halo size while the total energy tends to a constant. With this, we analyzed the possible scenarios of how this energy is distributed between a core and an envelope of the halo after virialization. This allowed us to formulate the predictions of our model for the final core-envelope structure, which are in agreement with the ULDM haloes, fitted to describe astrophysical observations \cite{Goldstein, walker2007velocity, 7fp7-klxf}.

{\it Acknowledgments}-- The work of E.V.G. and Y.R. was support by the Swiss National Science Foundation through the Ukrainian-Swiss Joint research project "Cosmic waltz of baryonic and ultralight dark matter: interaction and dynamical interplay" (grant No. IZURZ2{\_}224972). K.K. acknowledges funding by the Deutsche Forschungsgemeinschaft (DFG, German Research Foundation) under Germany’s Excellence Strategy—EXC 2123 QuantumFrontiers—390837967. The authors are grateful to Bohdan Hnatyk, Yurii Bidasyuk, Andrii Chaika, and Srikanth Nagesh for useful discussions.

\bibliography{main}

\end{document}